\documentclass[final]{svjour2}
\usepackage{ifpdf}
\usepackage[pdftex]{graphicx}
\usepackage{epstopdf}
\DeclareGraphicsExtensions{.eps, .pdf, .jpg, .tif}
\usepackage{rotating}
\usepackage{amscd}
\usepackage{amsmath}
\usepackage{bm}
\usepackage{amssymb}
\usepackage{mathptmx}
\usepackage[numbers]{natbib}
\makeatletter
\journalname{Journal of Low Temperature Physics}

\newcommand{\onehalf}{\frac{\mbox{\footnotesize 1}}{\mbox{\footnotesize 2}}}

\newcommand{\twothirds}{\frac{\mbox{\footnotesize 2}}{\mbox{\footnotesize 3}}}

\newcommand{\grad}{\mbox{\boldmath$\nabla$}}
\newcommand{\He}{$^3$He}
\newcommand{\Heb}{$^3$He-B}

\newcommand{\Heaero}{$^3$He-aerogel}

\newcommand{\sgn}{\mbox{sgn}}
\newcommand{\gammaS}{\gamma_{\mbox{\tiny S}}}
\newcommand{\gammaN}{\gamma_{\mbox{\tiny N}}}
\newcommand{\width}{\gamma_{\mbox{\tiny imp}}}
\newcommand{\Tcb}{T_{c_0}}

\newcommand{\mfp}{\ell_{\mbox{\tiny el}}}
\newcommand{\be}{\begin{equation}}
\newcommand{\ee}{\end{equation}}
\newcommand{\ber}{\begin{eqnarray}}
\newcommand{\eer}{\end{eqnarray}}

\newcommand{\whtauz}{{\widehat{\tau}_{\mbox{\tiny 3}}}}
\def\vDel{\boldsymbol{\upDelta}}
\def\vR{{\bf R}}

\def\vp{{\bf p}}
\def\hp{\hat\vp}

\def\vv{{\bf v}}
\def\vf{{\bf f}}

\def\whone{\widehat{1}}
\def\whg{\widehat{g}}

\def\mfg{\mathfrak{g}}
\def\whmfg{\widehat{\mathfrak{g}}}

\def\whu{\widehat{u}}
\def\wha{\widehat{a}}
\def\whb{\widehat{b}}
\def\whmft{\widehat{\mathfrak{t}}}
\def\vimp{\mbox{\footnotesize{V}}}
\def\whv{\widehat{\vimp}}
\def\whsigma{\widehat{\sigma}}
\def\whsigimp{\widehat{\sigma}_{\mbox{\tiny imp}}}
\def\whDel{\widehat{\Delta}}
\def\whDelt{\widehat{\widetilde{\Delta}}}
\def\tr#1{\mbox{\sf Tr}_{\mbox{\tiny 4}}\left\{#1\right\}}

\newcommand{\sumeps}{\pi T\sum_{\epsilon_n}}

\newcommand{\eps}{\epsilon_n}
\newcommand{\epst}{\widetilde{\epsilon}_n}
\begin{document}
\title{Thermodynamic Potential for \\ \hspace*{3ex} Superfluid \He\ in Aerogel}
\author{Sarosh Ali$^1$, Liangsheng Zhang$^{1,2}$ and \\ J. A. Sauls$^{1,\dag,\$}$}
\institute{1. Department of Physics \& Astronomy, Northwestern University, Evanston, Illinois, USA\\
           2. Department of Physics, Hong Kong University of Science and Technology,\\
           \hspace*{2ex}Clear Water Bay Road, Hong Kong\\
		   $\dag$. Corresponding author: \email{sauls@northwestern.edu} \\
		   \$. This work was supported by National Science Foundation Grant DMR-0805277.
		  }
\date{\today}
\maketitle
\begin{abstract}
We present a free energy functional for superfluid \He\ in the presence of homogeneously distributed
impurity disorder which extends the Ginzburg-Landau free energy functional to all temperatures. We
use the new free energy functional to calculate the thermodynamic potential, entropy, heat capacity
and density of states for the B-phase of superfluid \He\ in homogeneous, isotropic aerogel. 
\end{abstract}
\vspace*{-5mm}
\section*{Introduction}

Liquid \He\ infused into high porosity silica aerogel provides us with a model physical system for
studying the effects of quenched disorder on an Fermi liquid with unconventional pairing
\cite{hal04}. Indeed \Heaero\ provides us with a system where the effects of disorder can be
explored over the entire range from weak to strong disorder without modifying the basic interactions
of the Fermi liquid.

Theoretical analysis by Thuneberg et al. \cite{thu98} showed that the effects of scattering by a
homogeneous, isotropic medium enhanced the stability of the Balian-Werthamer (BW) phase relative to
the anisotropic ABM or Planar phase. These authors also calculated the reduction in the transition
temperature, $T_c$, the order parameter, $\whDel(\hp)$, and condensation energy,
$\Delta\Omega(p,T)$, in the Ginzburg-Landau (GL) limit.
Experimentally, the equilibrium phase of \Heaero\ is consistent with the Balian-Werthamer (BW)
state, modified by de-pairing, over the entire phase diagram, except perhaps in a narrow region of
temperatures near $T_c$ at high pressures \cite{spr96,bar00,sau05}. Measurements of the heat
capacity of \Heaero\ at intermediate and high pressures provide quantitative measurements of
de-pairing, including the reduction in $\Delta C/\gammaN \Tcb$ as well as indirect evidence of
gapless excitations with a finite density of states, $N(0)$, at the Fermi energy \cite{cho04a}.

We report results based on a free energy functional which extends the GL theory for superfluid
\He\ in the presence of homogeneously distributed impurity disorder to all temperatures. This
functional is obtained by a reduction of the Luttinger-Ward functional to leading order in the
expansion parameters of Fermi-liquid theory.
Results for the thermodynamic potential, entropy, heat capacity and density of states for the
B-phase of superfluid \Heaero\ in the weak-coupling limit are reported.
For impurity scattering in the unitary limit the B-phase exhibits gapless behavior over
the entire phase diagram for high-porosity aerogels with an elastic mean-free path
$\mfp=1500\,\mbox{\AA}$. 
\vspace*{-5mm}
\section*{Free Energy Functional}

Serene and Rainer (SR) obtained a functional of the low-energy, quasiclassical propagator,
$\whg(\hp,\eps)$, and self-energy, $\whsigma(\hp,\eps)$, for a superfluid Fermi liquid starting from
the Luttinger-Ward-DeDominicis-Martin (LWDM) functional \cite{lut60,ded64} of the exact propagator and self energy and subtracting the functional evaluated
at the stationarity condition corresponding to the normal Fermi liquid \cite{ser83}. This
subtraction confines the functional to the low-energy states near the Fermi surface.\footnote{We
follow the notation in Ref. \cite{ser83}, except that we denote $4\times 4$ matrices in Nambu space
by a widehat, e.g. $\whg$, while narrow hats refer to unit vectors, e.g. $\hp$, is the unit vector
in the direction of the Fermi momentum, $\vp_f = p_f\hp$.} Note that $\whsigma$, represents
corrections to the leading order normal-state self-energy. Similarly, the subtracted
$\Phi$-functional can be expressed as a functional of the low-energy propagator and effective
interactions represented by block vertices that sum to all orders the bare interaction and higher
order processes mediated by the high-energy propagator. So long as the block vertices do not 
introduce a new low-energy scale, then the contributions to the subtracted $\Phi$-functional can be
classified by their order in one or more small parameters of Fermi-liquid theory. The
SR functional has the form,
\ber\label{SR_functional}
\Delta\Omega\bigl[\whg,\whsigma\bigr] 
&=&
-\onehalf Sp^{\prime}\left\{\whsigma\whg\right\} 
+ 
\Delta\Omega_{\ln} 
+ 
\Phi\bigl[\whg\bigr]
\,,
\\
\Delta\Omega_{\ln}
&=&
-\onehalf Sp^{\prime}\Bigl\{\int d\xi_{\vp}
\left[
  \ln_{\otimes}\left(-i\epsilon_n\whtauz + \xi_{\vp} + \whsigma\right) 
- \ln_{\otimes}\left( -i\epsilon_n\whtauz + \xi_{\vp}\right)
\right]
\Bigr\}
\,,
\label{ln_functional}
\eer
where $Sp^{\prime}\{...\} \equiv 2N_f\int d^3R\,\int \frac{d\Omega_{\hp}}{4\pi}\,\sumeps
\tr{\ldots}$ and $ \tr{\ldots}$ is the trace in Nambu space. The SR functional is general enough to
describe inhomogeneous equilibrium states of a superfluid Fermi liquid which vary slowly on the
atomic scale. In such cases the convolution product attached to the $\ln$-functional is defined as
$\wha\otimes\whb(\vp,\vR)=\exp\Bigl\{
\frac{i}{2}\left(\grad_{\vR}^{a}\cdot\grad_{\vp}^{b}-\grad_{\vp}^{a}\cdot\grad_{\vR}^{b}\right)
\Bigr\}$
$\wha(\vp,\vR)\,\whb(\vp,\vR)$. For example $\xi_{\vp}\otimes\wha = (\xi_{\vp} - 
\frac{i}{2}\vv_f\cdot\grad_{\vR})\wha(\vp,\vR)$.
The SR functional is stationary with respect to variations of $\whg$ and $\whsigma$. The latter
condition generates,
\be\label{propagator}
\whg(\hp,\vR;\eps)=
\int d\xi_{\vp}\left(i\whtauz\eps - \xi_{\vp} - \whsigma\right)_{\otimes}^{-1}
\,,
\ee
which can be transformed {\`a} la Eilenberger into the quasiclassical transport equation
\cite{eil68}, 
$
\Bigl[i\whtauz\eps -\whsigma,\,\whg\Bigr] + i\vv_f\cdot\grad_{\vR}\whg = 0
$,
and normalization condition, $\whg(\hp,\vR;\eps)^2 = -\pi^2\whone$.
Stationarity of the SR functional with respect to variations of the propagator,
$\whg\rightarrow\whg+\delta\whg$, identifies the self energy with the skeleton expansion obtained
from the $\Phi$-functional,
\be
\delta\Phi\bigl[\whg\bigr]\equiv \Phi\bigl[\whg+\delta\whg\bigr]-\Phi\bigl[\whg\bigr] =
\onehalf Sp^{\prime}\{\whsigma\delta\whg\}
\,.
\ee
To leading order in the small expansion parameter the $\Phi$-functional is given by
\be
\Phi_{\mbox{\tiny ff}} 
= N_f\int d^3R\,
\int\frac{d\Omega_{\hp}}{4\pi}
\frac{d\Omega_{\hp'}}{4\pi}\,
v(\hp,\hp')\,
T\sum_{\eps}^{\prime}\vf(\hp,\vR;\eps)
\cdot
T\sum_{\eps'}^{\prime}\,\bar{\vf}(\hp',\vR;\eps')
\,,
\ee
where $v(\hp\cdot\hp')$ is the odd-parity, spin-triplet pairing interaction, while
$\vf(\hp,\vR;\eps)$ and $\bar{\vf}(\hp,\vR;\eps)$ are the spin-triplet components of the the
off-diagonal ``anomalous'' components of the propagator. This functional generates the mean-field
pairing self energy,
\be\label{f_self-consistency}
\vDel(\hp,\vR)
=\int\frac{d\Omega_{\hp'}}{4\pi}v(\hp,\hp')T\sum_{\eps'}^{\prime}\,\vf(\hp',\vR;\eps')
\,.
\ee
The Matsubara sums include a cutoff, $\Omega_c$, that restricts the pairing interaction to the
low-energy band near the Fermi level.
At this same order the diagonal mean-fields, i.e. the Landau molecular fields, also contribute.
These terms are important for states perturbed by external fields or for inhomogeneous phases, but
vanish for the homogeneous equilibrium state. For the applications discussed below the Landau
molecular field self energy is omitted.

In order to obtain a generalization of the GL functional 
we invert the self-consistency condition for the anomalous propagator to reduce the SR functional to
a functional of the order parameter, $\vDel$. Thus, $\Phi_{\mbox{\tiny ff}}$ can be evaluated to give,
\ber
\Phi_{\mbox{\tiny ff}} 
&=&  N_f \int d^3R\,
\frac{1}{v_{1}}\int\frac{d\Omega_{\hp}}{4\pi}\,|\vDel(\hp,\vR)|^2
\,,
\eer
for pure p-wave pairing. Similarly, in Eq. (\ref{SR_functional}), $-\onehalf
Sp^{\prime}\left\{\whsigma\whg\right\}=-2\Phi_{\mbox{\tiny ff}}$. 

In order to include the interaction of liquid \He\ with the structure of silica aerogel we 
explicitly include a one-body interaction, $\whv=\vimp\whone$, representing the interaction of
quasiparticles with the static impurity potential, $\vimp(\vp,\vp';\{\vR_i\})$.
The impurity potential enters via the $\ln$-functional, and the stationarity condition with respect
to $\whsigma$ generates Eq. (\ref{propagator}), but now with the impurity potential as an addition
to the self energy $\whsigma$. The impurity potential depends on the static configuration of the
impurities, $\{\vR_i\}$, which are treated as random variables governed by a probability
distribution, $P(\{\vR_i\})$. The impurity-averaged propagator is defined by averaging the
propagator for a specific configuration of impurities over the probability distribution for a
particular configuration,
\be
\whmfg(\hp,\vR;\eps) 
=
\int d\xi_{\vp}\,
\prod_{i=1}^{N_s}\int d^3R_i\,P(\{\vR_i\})\,
\left(i\eps\whtauz - \xi_{\vp} - \whDel - \whv(\{\vR_i\})\right)_{\otimes}^{-1}
\,.
\ee
The impurity average generates an additional self energy term describing the scattering of
quasiparticles and correlated pairs by the random potential. The derivation of the transport
equation carries through for the ensemble averaged propagator, which now takes the form
\be\label{transport_impurity-average}
\Bigl[i\whtauz\eps -\whDel - \whsigimp,\,\whmfg\Bigr] + i\vv_f\cdot\grad_{\vR}\whmfg = 0
\,.
\ee
For uncorrelated impurities the impurity self energy, to leading order in $1/\tau E_f$, is
determined by mean density of impurities and the self-consistent {\sf t}-matrix for impurity
scattering,
\ber\label{sigma_impurity}
\whsigimp(\hp,\vR;\eps) 
&=& n_{\mbox{\tiny imp}}\,\whmft(\hp,\hp,\vR;\eps)
\\
\whmft(\hp,\hp',\vR;\eps) 
&=& \whu(\hp,\hp')
+ N_f\int\frac{d\Omega_{\hp''}}{4\pi}\,
\whu(\hp,\hp'')\,\whmfg(\hp'',\vR;\eps)\,\whmft(\hp'',\hp',\vR;\eps)
\,.\quad
\eer
For the results presented here we consider isotropic impurities with dominant scattering in the
s-wave channel, i.e. $\whu = u_0\whone$. In this case the {\sf t}-matrix reduces to,
\be\label{tmatrix_s-wave}
\whmft(\vR;\eps) = u_0
\left(\whone + N_f \langle\whmfg(\hp,\vR;\eps)\rangle_{\hp}\,\whmft(\vR;\eps)\right)
\,,
\ee
where $\langle\whmfg(\hp,\vR;\eps)\rangle_{\hp}$ is the average of the propagator over the Fermi
surface. 
For normal \He\ the propagator reduces to $\whmfg_{\mbox{\tiny N}}=-i\pi\whtauz\sgn(\eps)$, and the
normal-state {\sf t}-matrix is parameterized by the s-wave scattering phase shift,
$\delta_0=\tan^{-1}(\pi N_f\,u_0)$, and reduces to $\whmft_{\mbox{\tiny N}}(\eps) = \frac{1}{\pi
N_f}\,\sin\delta_0\,e^{-is_{\epsilon}\delta_0\whtauz}$, where $s_{\epsilon}=\sgn({\epsilon_n})$. In
this minimal model for aerogel scattering, the mean density of impurities $n_{\text{\tiny s}} =
1/\sigma\ell$, and the scattering rate, $1/\tau$, for normal-state quasiparticles are fixed by the
mean free path, $\ell=v_f\tau$, and scattering cross-section, $\sigma=4\pi\lambda_f^2\,\bar\sigma_0$
where $\lambda_f = \hbar/p_f$ is the Fermi wavelength and $\bar\sigma_0 = \sin\delta_0^2$ is the
dimensionless cross-section \cite{buc81}.

In order to average the free energy functional in Eq. (\ref{SR_functional}) over the impurity
ensemble we must average the $\ln$-functional in Eq. (\ref{ln_functional}). This is facilitated by
the integral representation for the $\ln$-functional,
\ber
&-&
\tr{\ln_{\otimes}\left(-i\eps\whtauz+\xi_{\hp}+\whDel(\vp,\vR)+\whv(\{\vR_i\})\right)}
\nonumber\\
&=&
\tr{
i\whtauz
\int_{\eps}^{\infty}d\lambda\,
\left(i\lambda\whtauz-\xi_{\vp}-\whDel(\hp,\vR)-\whv(\{\vR_i\})\right)_{\otimes}^{-1}
}
\label{integral_representation}
\,.\quad
\eer

For the subtracted functional we can now carry out the ensemble average, and interchange the
$\xi_{\vp}$ integration with the auxiliary integration over the Matsubara energy to obtain the
ensemble averaged free energy functional,
\ber\label{free-energy_functional_impurity-averaged}
\Delta\overline{\Omega}[\whDel] 
=  N_f\int d^3R\,\int\frac{d\Omega_{\hp}}{4\pi}\,
\Bigg\{
&-&\frac{1}{v_1}\,|\vDel(\hp,\vR)|^2
\\
&+&
\onehalf 
T\sum_{\eps}^{\prime}\int_{\eps}^{\infty}d\lambda
\tr{i\whtauz[\whmfg(\hp,\vR;\lambda) - \whmfg_{\text{\tiny N}}(\lambda)]}
\Bigg\}
\,,
\nonumber
\eer
where $\whmfg_{\mbox{\tiny N}}(\lambda)=-i\pi\whtauz\sgn(\lambda)$ and 
$\whmfg(\hp,\vR;\lambda)$ is obtained from the solution to the Eqs.
(\ref{transport_impurity-average},\ref{sigma_impurity} and \ref{tmatrix_s-wave}) for fixed
$\whDel(\hp,\vR)$ and auxillary Matsubara energy, $i\lambda$. 
This free energy functional can be used to evaluate the thermodynamic potential for a broad range of
equilibrium states of unconventional superconductors in the presence of impurity disorder; it also
represents an extension of the Eilenberger's free energy functional for conventional superconducting
alloys in the Born impurity scattering limit \cite{eil66}. As an application we calculate the
thermodynamic potential for the BW state in the presence of homogeneous, isotropic impurity
scattering as a model for the B-phase of \He\ in silica aerogel.
%

For spatially uniform states the solution to the transport equation and normalization condition 
for the propagator can be expressed as
\be\label{propagator_renormalized}
\whmfg = -\pi\frac{i\epst\whtauz - \whDelt(\hp;\eps)}
				  {\sqrt{\epst^2 + |\widetilde{\vDel}(\hp;\eps)|^2}}
\,,
\ee
where $\epst=\eps+i n_{\mbox{\tiny imp}}\,\mathfrak{t}_{\mbox{\tiny 3}}(\eps)$ and
$\widetilde{\vDel}=\vDel(\hp) + \vDel_{\mbox{\tiny imp}}(\eps)$ are the Matsubara energy and order
parameter renormalized by the impurity self-energy. However, $\vDel_{\mbox{\tiny imp}}(\eps)$
vanishes for pure s-wave scattering because the order parameter vanishes when averaged over the
Fermi surface, i.e. $\langle\vDel(\hp)\rangle_{\hp} = 0$ \cite{cho89b}. Thus, we obtain from the
$\whtauz$-component of the {\sf t}-matrix,
\be
\epst = \eps + \frac{1}{2\tau}
\Bigg\langle
\frac{\epst\sqrt{\epst^2+|\vDel(\hp)|^2}}{\epst^2+|\vDel(\hp)|^2(1-\bar\sigma_0)}
\Bigg\rangle_{\hp}
\,.
\ee

The self-consistency equation for the order parameter is obtained by evaluating Eq.
(\ref{f_self-consistency}) with the solution for $\vf(\hp;\eps)$ obtained from Eq.
(\ref{propagator_renormalized}). For unitary spin-triplet states ($\vDel\times\vDel^*=0$) the gap
equation becomes,
\be\label{gap_equation_renormalized}
\vDel(\hp)= -
\int\frac{d\Omega_{\hp'}}{4\pi}v(\hp,\hp')
\pi T\sum_{\eps}^{\prime}\frac{\vDel(\hp')}{\sqrt{\epst^2+|\vDel(\hp')|^2}}
\,.
\ee
For pure \He\ the linearized gap equation (LGE) with $\vDel\rightarrow 0$ and $1/2\tau\rightarrow 0$
determines the bulk transition temperature, $T_{c_0}$, in terms of the p-wave pairing interaction,
$v_1$, and the bandwidth of low-energy states, $\Omega_c$, i.e. $\mathsf{K}(T_{c_0})=-1/v_1$, with
$\mathsf{K}(T)=2\pi T\sum_{\eps\ge 0}^{\prime}(1/\eps) \simeq\ln(1.13\Omega_c/T)$. The LGE
also regulates the Matsubara sum and removes the pairing interaction and cutoff in favor of the pure
\He\ transition temperature, $T_{c_0}$.
\be\label{gap_equation_regulated}
\ln\left(\frac{T}{T_{c_0}}\right) 
\vDel(\hp)
= 
2\pi T\sum_{n\ge 0}
	\left(
	\frac{1}{\sqrt{\epst^2 + |\vDel(\hp)|^2}}
	-
	\frac{1}{\eps}
	\right)
\vDel(\hp)\,
\,.
\ee

The renormalization of the Matsubara energy encodes the effects of de-pairing by the scattering of
pair-correlated quasiparticles off the distribution of impurities. This is evident in suppression of
the superfluid transition and the order parameter.
The suppression of the superconducting transition is given by the solution of Eq.
(\ref{gap_equation_regulated}) for $|\vDel|\rightarrow 0^+$. In this limit, 
$\epst\rightarrow\eps+\frac{1}{2\tau}\sgn(\eps)$ and $T\rightarrow T_c$,
\ber\label{Tc_suppression}
\ln\left(\frac{T_c}{T_{c_0}}\right) 
&=& 
\sum_{n\ge 0}
		\left(
		\frac{1}{n+\onehalf + \onehalf\,\frac{1}{2\pi\tau T_c}}
	-
	\frac{1}{n+\onehalf}
	\right)
\,,
\eer
and we obtain the Abrikosov-Gorkov formula \cite{thu98}. Note that the pair-breaking parameter,
$\alpha=1/2\pi\tau T_{c_0} = \xi_0(p)/\ell$, where $\xi_0=\hbar v_f/2\pi k_B T_{c_0}$ is the pure
\He\ coherence length and $\ell=v_f\tau$ is identified with the transport mean free path. The full
range from weak to strong pair breaking can be explored by varying the pressure and aerogel density.
The critical point in the phase diagram where $T_c = 0$ is given by $\alpha_c \simeq 0.281$. For
98\% aerogel with a mean free path of $\ell\simeq 1400\,\mbox{\AA}$ this gives a critical pressure
of $p_c\simeq 6\,\mbox{bar}$, below which the superfluid transition is completely suppressed
\cite{mat97}. For superfluidity to survive down to $p=0\,\mbox{bar}$ the mfp must exceed $\approx
3000\,\mbox{\AA}$, while for $\ell\lesssim 650\,\mbox{\AA}$ superfluidity is completely suppressed
for all pressures.

\begin{figure}[h!]
\begin{center}
\includegraphics[width=0.70\linewidth,keepaspectratio]{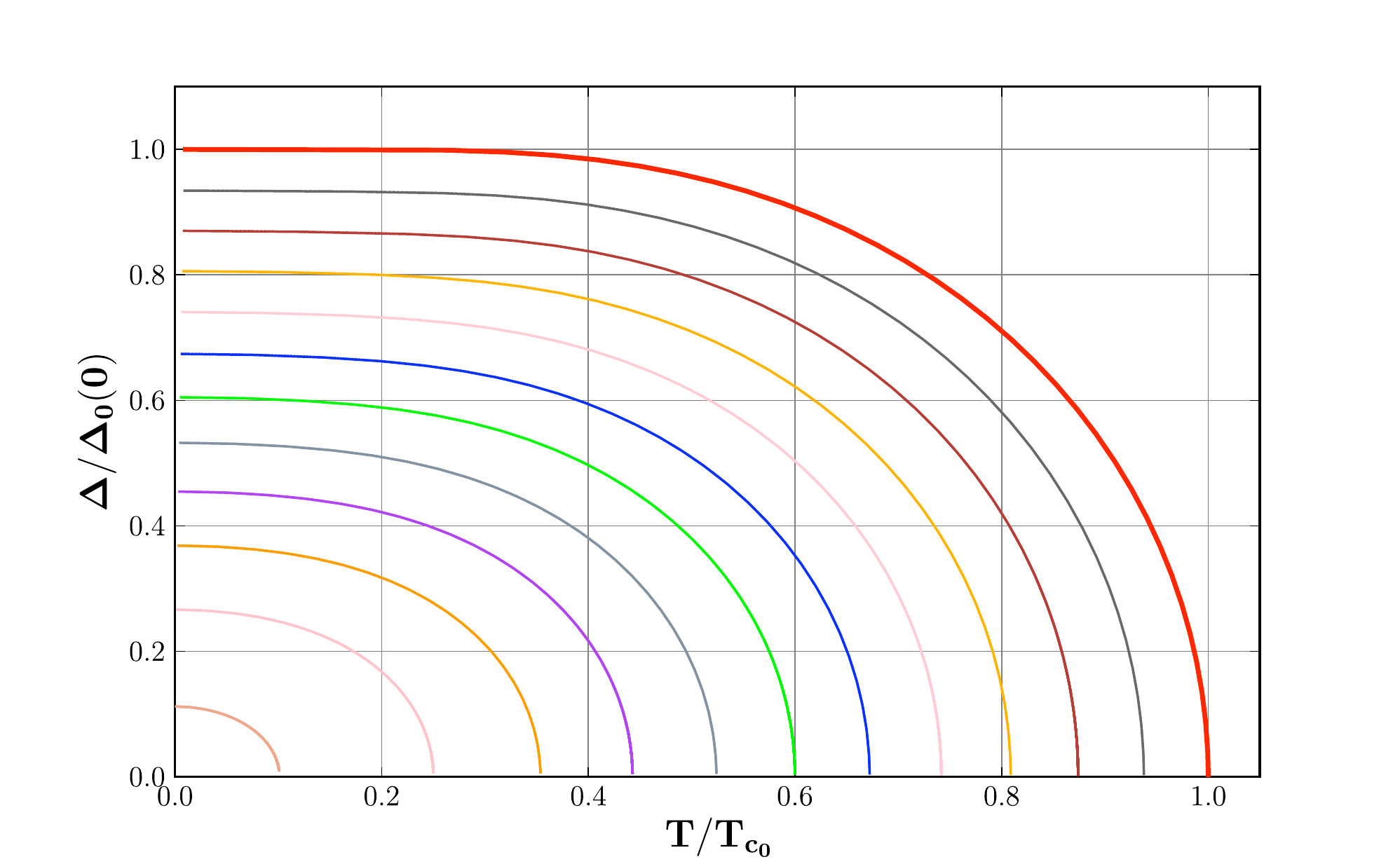}
\end{center}
\caption{(Color online) Suppression of the B-phase order parameter in the unitary limit as a
function of $\alpha=1/2\pi\tau T_{c_0}$. $\Delta(T)$ is normalized by $\Delta_0(0)=1.78 k_B T_{c_0}$
for pure \Heb\ in the weak-coupling limit (red curve). The curves for $0 < \alpha\le 0.275$ are
shown for $\alpha$ in steps of $0.025$.}
\label{figure:gap}
\end{figure}

For the B-phase, $|\vDel(\hp)|^2=|\Delta|^2$ is isotropic, the angular integration over the Fermi
surface is trivial and we can solve the gap equation numerically to obtain the magnitude of the
order parameter.
Figure \ref{figure:gap} shows the de-pairing effect on both the transition temperature and the
magnitude of the order parameter over the full temperature range as a function of $\alpha$ for
unitary scattering.

The free energy functional for spatially uniform phases of \Heaero\ is obtained by carrying out the
integration in Eq. (\ref{free-energy_functional_impurity-averaged}) over the impurity renormalized
propagator in Eq. (\ref{propagator_renormalized}). For unitary spin-triplet states
and impurity scattering in the unitary scattering limit we obtain,
\ber\label{Omega_functional_impurity}
\Delta\bar{\Omega}\bigl[\vDel,\vDel^*\bigr] 
&=& N_f V\,\Bigg\langle
-\frac{1}{v_1}\,|\vDel(\hp)|^2
+
2\pi T\sum_{\eps}^{\prime}
\Big[
\left(|\epst|-\sqrt{\epst^2+|\vDel(\hp)|^2}\right)
\nonumber\\
&&\qquad
\times\left(1+\frac{1}{2\tau|\epst|}\right)
+
\frac{1}{2\tau}\ln\left(\frac{\sqrt{\epst^2+|\vDel(\hp)|^2}}{|\epst|}\right)
\Big]
\Bigg\rangle_{\hp}
\,,
\\
\epst &=& \eps +
\Bigg\langle\frac{1}{2\tau}\frac{\sqrt{\epst^2+|\vDel(\hp)|^2}}{\epst}\Bigg\rangle_{\hp}
\,.
\label{renormalized_Matsubara}
\eer
This functional reduces to the weak-coupling functional for pure \He\ (Eq. 5.16 of Ref.
\cite{ser83}) for $\tau\rightarrow\infty$. One can also verify that the stationarity condition for
the ensemble averaged functional, $\delta\Delta\bar\Omega/\delta\vDel(\hp)^* = 0$, generates the
impurity renormalized gap equation, i.e. Eq. (\ref{gap_equation_renormalized}).

The thermodynamic potential, entropy and heat capacity of BW phase with homogeneous,
isotropic disorder are obtained by evaluating the ensemble averaged free energy functional with the
self-consistently determined solutions of Eqs. (\ref{gap_equation_regulated}),
(\ref{Omega_functional_impurity}) and (\ref{renormalized_Matsubara}). 

\begin{figure}[h!]
\begin{tabular}{ll}
\hspace*{-1cm}\includegraphics[width=0.60\linewidth,keepaspectratio]{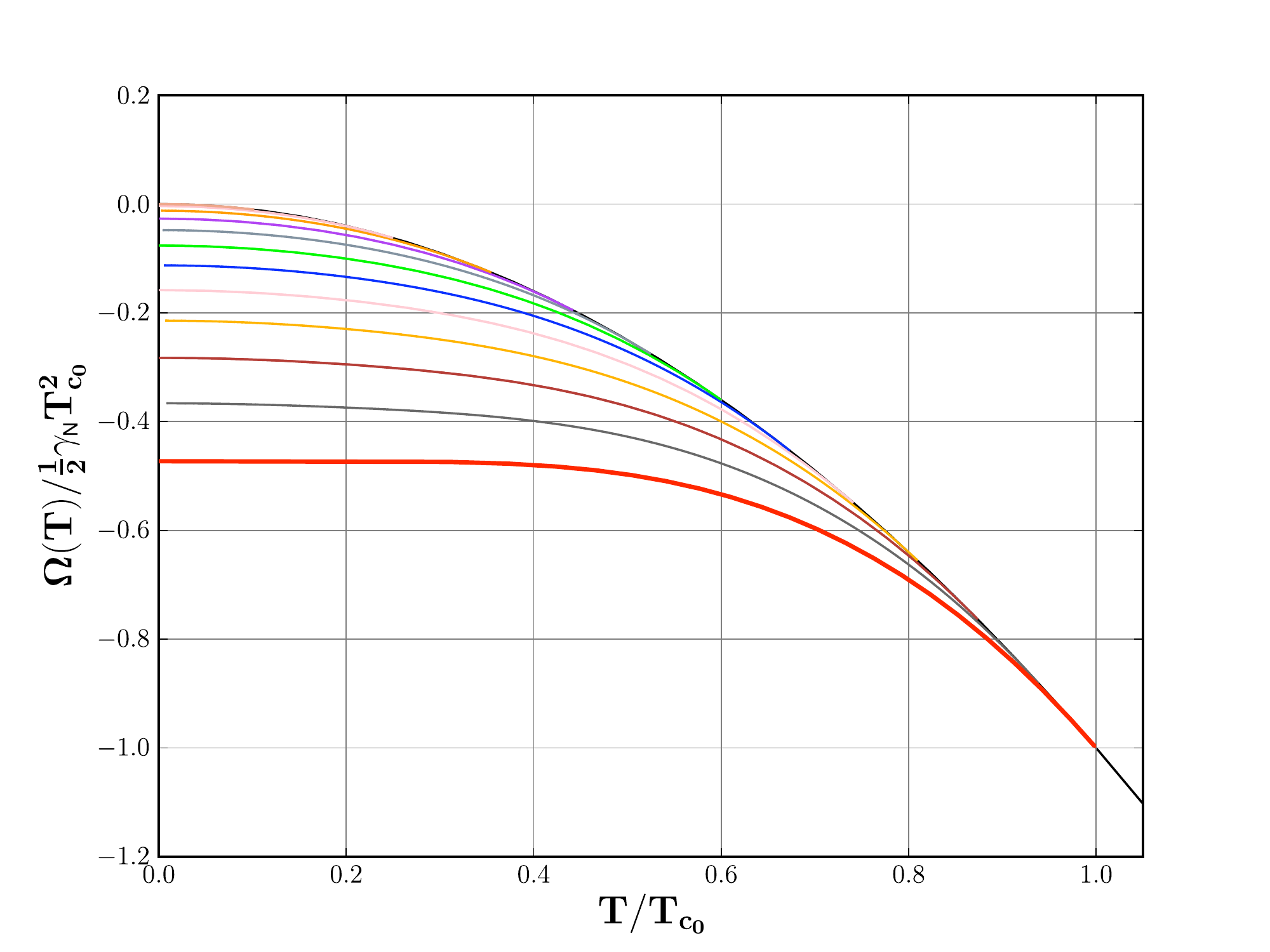}
&
\hspace*{-1.05cm}\includegraphics[width=0.60\linewidth,keepaspectratio]{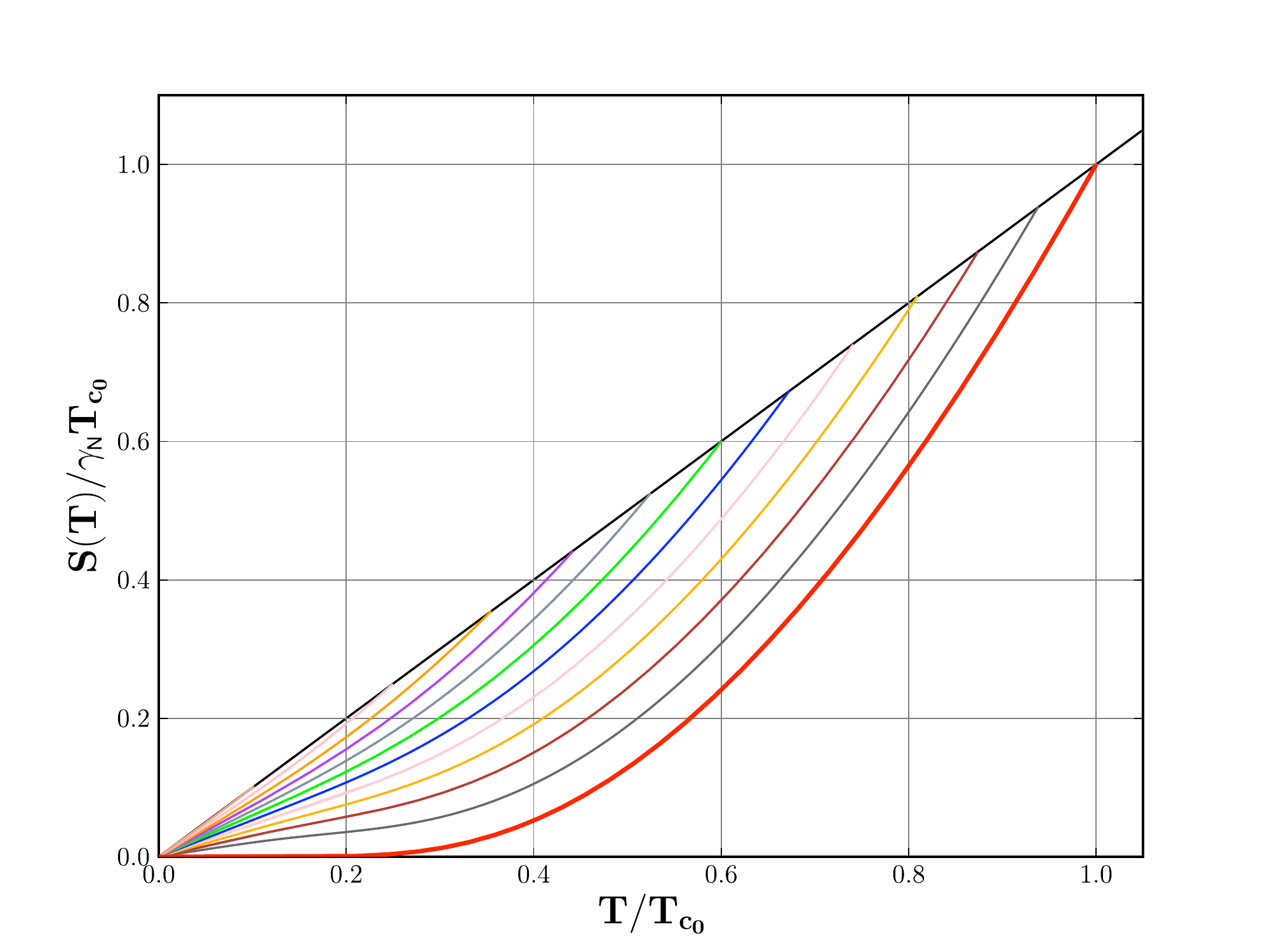}
\end{tabular}
\caption{(Color online) Thermodynamic potential (left panel) and entropy (right panel) for the
B-phase of \Heaero\ in the unitary limit as a function of $T/T_{c_0}$ for pair breaking ranging from
$0 \le \alpha\le 0.275$ in steps of $0.025$. Results for pure \Heb\ are shown as red curves.  
}
\label{figure:Omega}
\end{figure}
\vspace*{-6mm}
\section*{Thermodynamic Functions}

Figure \ref{figure:Omega} shows the thermodynamic potential (left panel) and entropy (right panel)
for the dirty B-phase for values of the pair breaking parameter spanning the range from weak to
strong pair breaking. Note that we plot the full thermodynamic potential, $\Omega_{\mbox{\tiny
S}}(T)=\Omega_{\mbox{\tiny N}}(T)+\Delta\bar\Omega(T)$, which includes the normal-state contribution
$\Omega_{\mbox{\tiny N}}(T)=-\onehalf\gammaN\,T^2$, and we normalized the potential in units of
$\onehalf\gammaN\,T_{c_0}^2$, where $\gammaN=\twothirds\pi^2 N_f$ is the Sommerfeld coefficient for
the normal state. The entropy was computed by numerically differentiating the thermodynamic
potential, $S(T)=-\partial\Omega/\partial T|_{\mbox{\tiny V}}$. For pure \Heb\ the entropy vanishes
exponentially for $T\ll T_{c_0}$, but for the disordered B-phase in the unitary limit the entropy
vanishes linearly for sufficiently low temperatures, $S(T)\rightarrow\gammaS T$, where $\gammaS$ is
the Sommerfeld coefficient for the gapless B-phase. The gapless regime is achieved for $T\ll\width$,
where $\width\simeq\sqrt{\Delta(0)/2\tau}$ is the bandwidth of the impurity-induced excitation
spectrum near the Fermi level.

\begin{figure} 
\begin{tabular}{ll}
\hspace*{-1cm}\includegraphics[width=0.60\linewidth,keepaspectratio]{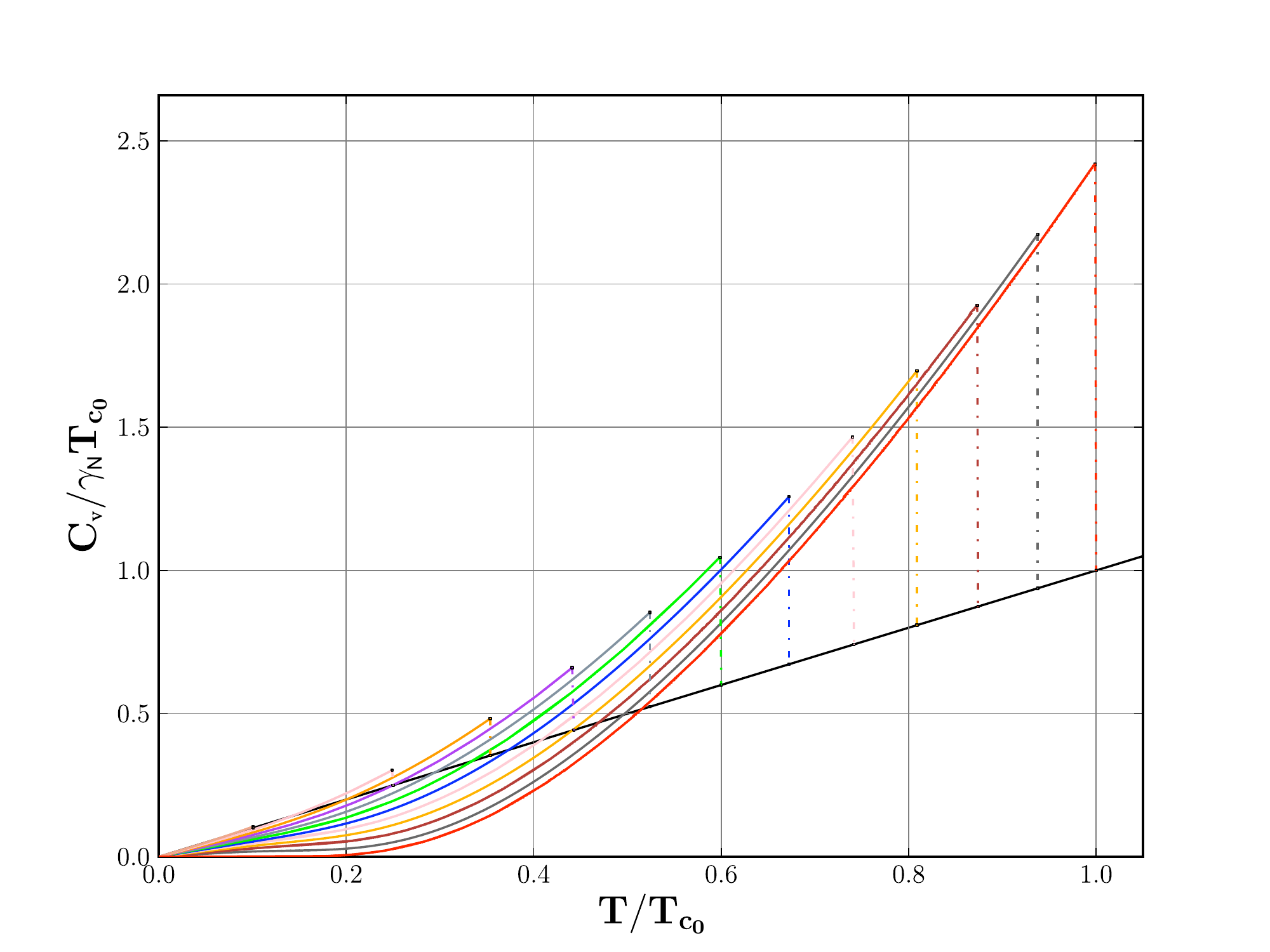}
&
\hspace*{-1.05cm}\includegraphics[width=0.60\linewidth,keepaspectratio]{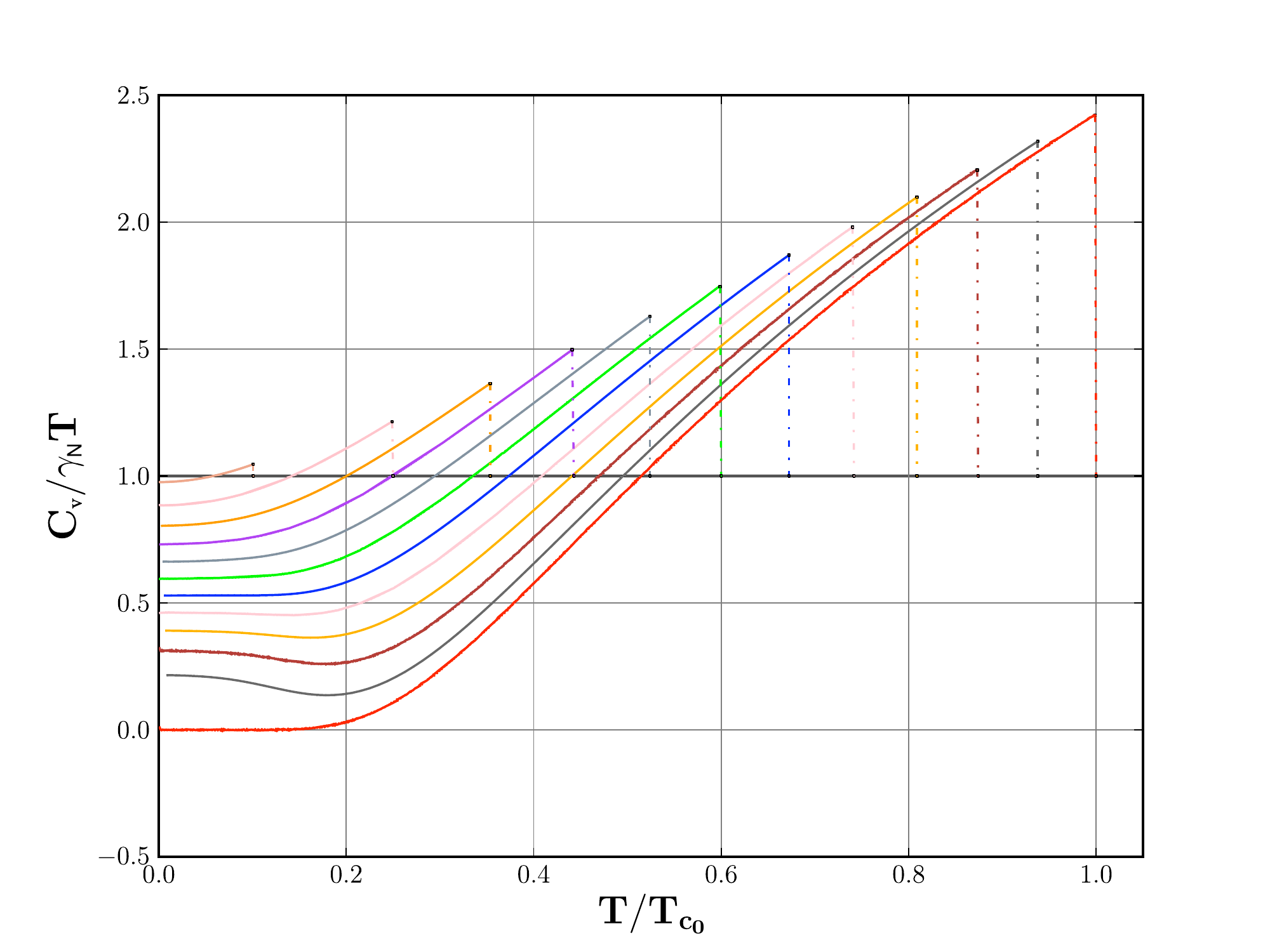}
\end{tabular}
\caption{(Color online) Specific heat capacity normalized by $\gammaN T_{c_0}$ (left panel) and by
$\gammaN T$ (right panel) for the B-phase of \Heaero\ in the unitary limit as a function of
$T/T_{c_0}$ for pair breaking ranging from $0 \le \alpha\le 0.275$ in steps of $0.025$. Results for
pure \Heb\ are shown as red curves.
}
\label{figure:Specific_Heat}
\end{figure}

The heat capacity provides both a quantitative measure of the order that develops at $T_c$, as well
as the spectrum of gapless states near the Fermi level. Figure \ref{figure:Specific_Heat} (left
panel) shows results for the specific heat, $C_{\mbox{\tiny V}}=T\partial S/\partial T|_{\mbox{\tiny
V}}$ over the full temperature range. The reduction of the heat capacity jump at $T_c$ computed from
the thermodynamic potential agrees with the value calculated from the GL theory of Ref. \cite{thu98}
to three significant figures for all values of the pair breaking
parameter. The right panel is the same data for the heat capacity plotted as $C_{\mbox{\tiny
V}}/\gammaN T$ vs. $T/T_{c_0}$.
For weak to modest pair-breaking, $\alpha=1/2\pi\tau\Tcb < 0.1$, corresponding to $\mfp >
1600\,\mbox{\AA}$ at $p=33\,\mbox{bar}$, $C_{\mbox{\tiny V}}/T$ is a non-monotonic function of $T$,
however the limiting low-temperature heat capacity is linear in temperature, $C_{\mbox{\tiny
V}}\rightarrow\gammaS T$, where $\gammaS$ is the Sommerfeld coefficient for temperatures below the
impurity bandwidth, $T\lesssim\width$. The non-monotonic behavior of $C_{\mbox{\tiny V}}/T$ for
$\alpha\lesssim 0.1$ results from the gap between the impurity band and the continuum states with
$\varepsilon\ge|\Delta|$ (see Fig. \ref{figure:DOS}).
The limiting value provides the Sommerfeld coefficient for the gapless B-phase,
$\gammaS=\gammaN\lim_{T\rightarrow 0}C_{\mbox{\tiny V}}/\gammaN T$.
One can infer from the calculated Sommerfeld coefficient for the gapless B-phase the density of
states at the Fermi level, i.e. $\gammaS=\twothirds\pi^2\,N(0)$.

\section*{Density of States}

\begin{figure}[h!]
\begin{tabular}{ll}
\hspace*{-0.75cm}\includegraphics[width=0.525\linewidth,keepaspectratio]{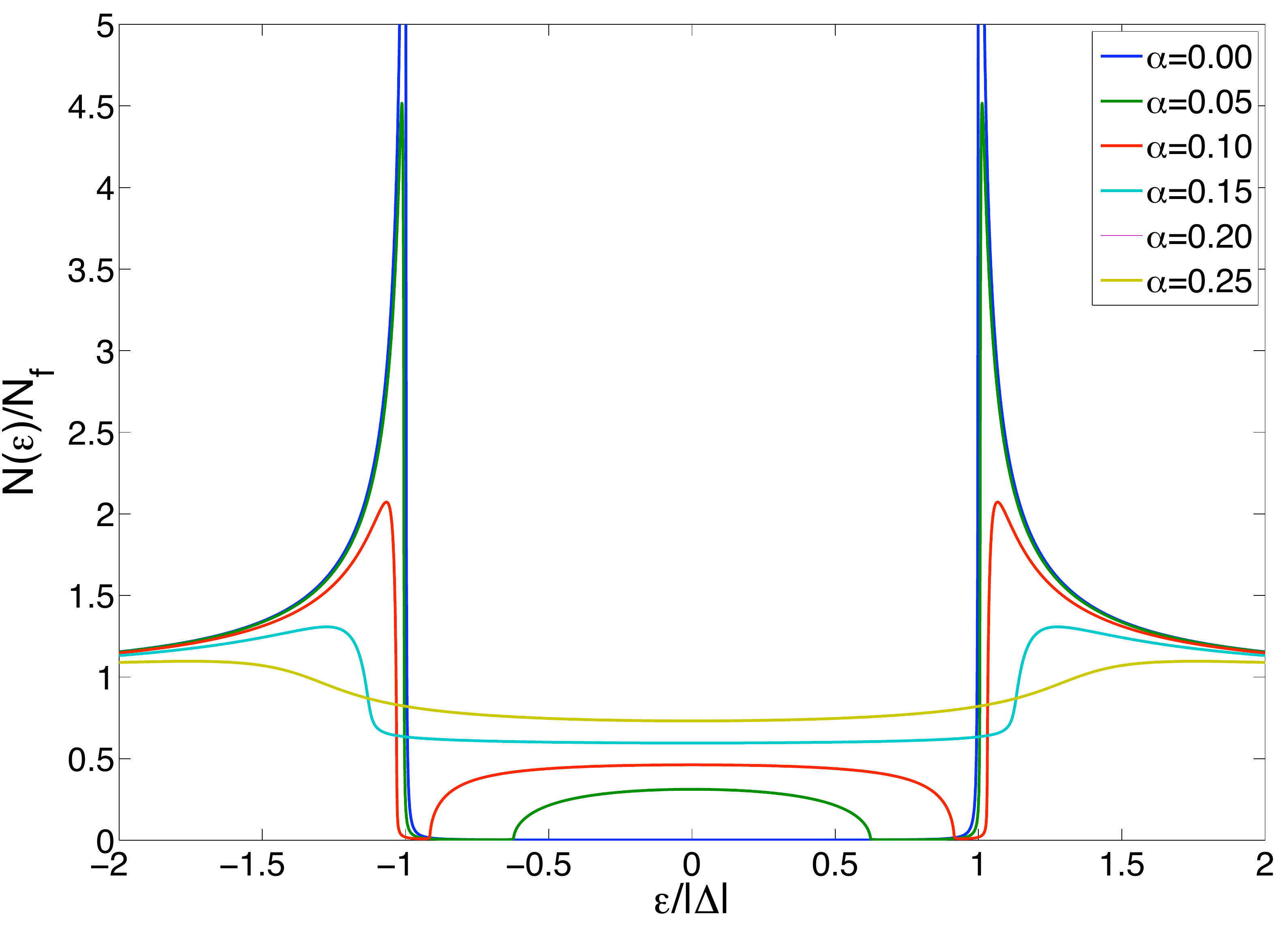}
&
\hspace*{-0.05cm}\includegraphics[width=0.5\linewidth,keepaspectratio]{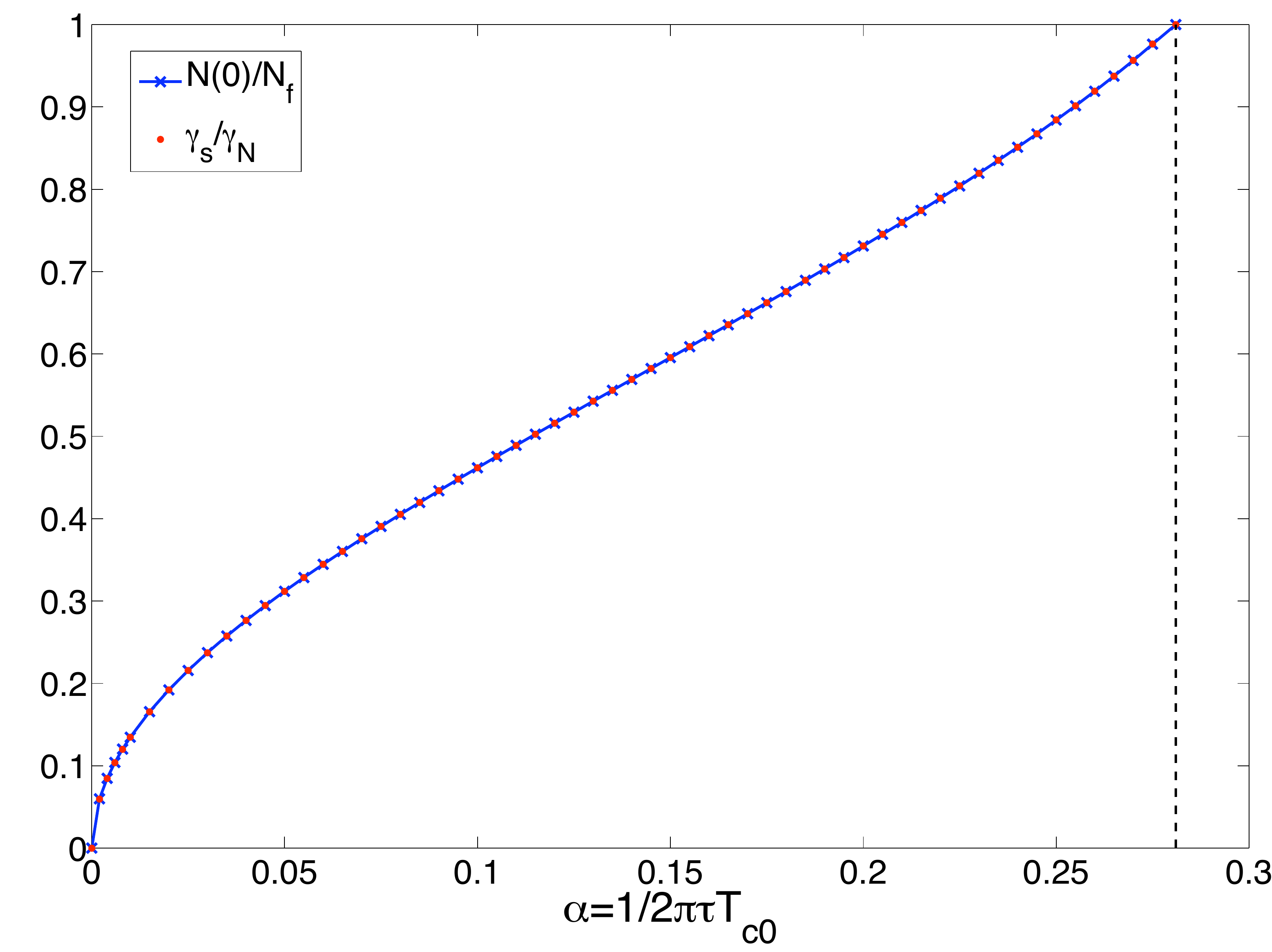}
\end{tabular}
\caption{(Color online) Density of states as a function of excitation energy from weak to strong
pair breaking (left). Comparison of the normalized density of states at the Fermi level (right
panel), $N(0)/N_f$ obtained by direct calculation from Eq. (\ref{DOS}), and the Sommerfeld ratio,
$\gammaS/\gammaN$ computed from the limit, $\lim_{T\rightarrow 0}C_{\mbox{\tiny S}}/T$.
}
\label{figure:DOS}
\end{figure}

On the other hand, the density of states can be calculated directly by analytic continuation of the
Matsubara propagator to real energies,
\ber 
\mfg(i\eps)
&\xrightarrow[]{i\eps\rightarrow\varepsilon+i0^+}&
\mfg^R(\varepsilon)
=
-\pi\frac{\widetilde{\varepsilon}^R(\varepsilon)}
         {\sqrt{|\Delta|^2-(\widetilde{\varepsilon}^R(\varepsilon))^2}}
\,,
\label{gR}
\eer
\ber
\widetilde{\epsilon}^R
&=&
\epsilon+i0^+ 
-\frac{1}{2\tau}\frac{\sqrt{|\Delta|^2-(\widetilde{\epsilon}^R(\varepsilon))^2}}
                     {\widetilde{\epsilon}^R}
\,.
\label{epstR}
\eer
\be 
N(\varepsilon)
=N_f\,\Bigg(-\frac{1}{\pi}\mathrm{Im}\,\mfg^R(\varepsilon)\Bigg)
\,,
\label{DOS}
\ee
where $N(\varepsilon)$ is the density of states (DOS) at excitation energy $\varepsilon$ measured
relative to the Fermi level. Figure \ref{figure:DOS} shows the results for the DOS (left panel) from
weak to strong pair breaking calculated in the unitary limit for $T\rightarrow 0$. Note that the
excitation energy for each value of pairbreaking is normalized by the appropriate value of the order
parameter, $\Delta(0)$.
The divergence of $N(\varepsilon)$ at $\varepsilon=\pm|\Delta|$ is suppressed by impurity
scattering, and gapless excitations are created near the Fermi level for any $\alpha\ne 0$ 
in the unitary scattering limit.

The normalized density of states at $\epsilon=0$ can be compared with that inferred from the
calculated from the Sommerfeld coefficient. This provides a strong test of our calculations for the
thermodynamic potential, entropy and heat capacity, as well as the validity of the ensemble averaged
free energy functional obtained in Eq. (\ref{Omega_functional_impurity}). As Figure \ref{figure:DOS}
shows the results for $N(0)/N_f$ and $\gammaS/\gammaN$ (right panel) are in perfect agreement. Note
that $N(0)/N_f$ and $\gammaS/\gammaN$ vanish for $\alpha=0$ and both approach 1 continuously for
$\alpha\rightarrow\alpha_c\approx 0.281$.

In summary we have obtained an ensemble averaged free energy functional for the superfluid \He\ in
in the presence of homogeneous, isotropic impurity disorder. This functional is based on a
quasiclassical reduction of the Luttinger-Ward functional. Results for the thermodynamic potential,
entropy and heat capacity were obtained by numerical evaluation of the stationarity conditions. The
predicted Sommerfeld ratio based on the new functional agrees exactly with a direct calculation of
the DOS from a solution of the quasiclassical transport equation.
A quantitative comparison between theory and experimental measurements of the heat capacity of
superfluid \Heb\ in aerogel \cite{cho04a} made at intermediate and high pressures will be 
discussed in a separate report.

%

\end{document}